\documentclass[prb,superscriptaddress,showpacs]{revtex4-1}
\makeindex
\usepackage{graphicx}
\usepackage[utf8]{inputenc}
\usepackage{dcolumn}
\usepackage{bm}
\usepackage{amsmath}
\usepackage{amssymb}
\usepackage{ulem}
\usepackage{color}
\usepackage{float}

\def\X{{\mathcal X}}

\def\H{{\mathcal H}}

\def\Z{{\mathcal Z}}
\def\CV{C_V}
\def\EG{e_{h,0}}
\def\se{{\rm s}}
\def\Tr{{\rm Tr}}
\def\HT{{\rm HT}}

\def\PA{{\rm PA}}
\def\SE{{\rm SE}}

\begin{document}

\title{Spin susceptibility of  quantum magnets from high to low temperatures. }
\author{B. Bernu}
   \affiliation{LPTMC, UMR 7600 of CNRS, UPMC, Paris-Sorbonne, F-75252 Paris Cedex 05, France}
\author{C. Lhuillier}
   \affiliation{LPTMC, UMR 7600 of CNRS, UPMC, Paris-Sorbonne, F-75252 Paris Cedex 05, France}
\date{\today}
\begin{abstract}
We explain how and why all thermodynamic properties of spin systems can be computed in one and two dimensions in the whole range of temperatures overcoming the divergence towards zero temperature of the standard high temperature series expansions (HTE).
The method relies on an approximation of the entropy versus energy (microcanonical potential function) on the whole range of energies. The success is related to the intrinsic physical constraints on the entropy function and a careful treatment of the boundary behaviors.
This method is benchmarked against  two one-dimensional solvable models: the Ising model in longitudinal  field and the XY model in a transverse field. With ten terms in the HTE, we find a spin susceptibility within a few \% of the exact results in the whole range of temperature. The method is then applied to two two-dimensional models: the supposed-to-be gapped Heisenberg model and the $J_1$-$J_2$-$J_d$ model on the kagome lattice.
\end{abstract}

\maketitle

Recent years have seen an outburst of magnetic materials that might be realistic candidates for the long searched  spin liquids \cite{Shimizu2003,mendels2007,helton07,Yamashita2008,Yamashita2011,faak2012,Han2012,Han2012a}.  In this quickly maturing field, it is now highly desirable  to compare the experimental properties of these  new states of matter with theory. Modelization of the magnetic interactions in a Mott insulator is a challenge.
First principle calculations of the magnetic interactions are delicate\cite{Janson_2008,Jeschke2013}.  In a pragmatic approach, the experimentally measured specific heat $C_V$ and/or uniform spin susceptibility $\X$ can be compared against high temperature series expansions (HTE) of spin models.\cite{lohmann2014}
This simple trail is not sufficient for frustrated magnets.
In fact HTE diverge at low temperature and increasing the length of the series and/or using Pad\'e approximants do not help much to reach useful information at temperatures lower than the main interaction.
But, as was noticed at the early years of this quest \cite{r00}, the  interesting physics in frustrated systems appear in a range of temperatures at least an order of magnitude lower than the main coupling, and specially for competing interactions where the temperature  range available with the raw HTE is quite insufficient.

 Many mathematical methods have been tried to reach low temperature properties of magnets. For example, biased differential approximants have been successful to account for the N\'eel ground states of the Heisenberg  model on the square and the triangular lattice.\cite{Roger1998}
For spin liquids presently under investigation other tools are needed.
In refs.\cite{Bernu2001b,Misguich2005a} a different approach, based on the use of  sum rules,   was proposed to compute specific heat at zero magnetic field.
 Unfortunately in real materials, phonon and magnetic contributions to the specific heat are often mixed and extracting the magnetic contribution is delicate.
On the contrary, the experimental information on the magnetic susceptibility obtained by squid or NMR measurements is free of these uncertainties and it would  be  extremely valuable to have a way to use it.
The extension of refs.\cite{Bernu2001b,Misguich2005a} to spin-susceptibility calculation did not seem a priori possible
as its success was thought to be related to the existence of sum rules constraining the specific heat, sum rules which do not have equivalent for the spin-susceptibility.

In fact  deep physical reasons imply that refs.\cite{Bernu2001b,Misguich2005a} regularization and interpolation procedure is more powerful than  expected.   From a conceptual point of view the first key point is the move from an expansion of the free energy $f$ as a function  of the temperature $T$, to a expansion of the entropy $s$ as a function of the internal energy $e$. Elementary statistical mechanics tells us that these two descriptions (canonical versus microcanonical ensembles) are indeed equivalent and that all thermodynamic quantities can be computed at the thermodynamic limit in any of them.
The major drawback of the standard use of truncated HTE is the intrinsic divergence arising in the low temperature free energy:  trying to extend its range of validity towards $T=0$ is thus extremely difficult.
The choice  to build a reasonable approximation of  the entropy versus energy $s(e)$ (and by extension of $s(e,h)$, where $h$ is the external magnetic field) is more efficient because:
 $i)$ $s(e)$ is defined on the finite interval  from the ground state energy $e_0$, to $e_\infty$ (the average energy reached by the system for $T\rightarrow\infty$), and its boundary values are known: $s(e_0) = 0$ and $s(e_\infty) = \ln (2S+1)$.
 $ii)$ The series expansion of $s(e)$ at $e_\infty$  can be exactly deduced from the knowledge of the free energy HTE. 
$iii)$ In the absence of phase transition (one- or two-dimensional behavior) the function $s(e) $ is an infinitely derivable function on $]e_0,e_\infty]$, monotonously increasing ($s'(e)=1/T>0$) and concave (second derivatives of $s(e,h)$ negative because of  stability conditions of the thermodynamical equilibrium)
$iv)$ The  correct behavior of $s'(e)$ at $e_0$  can be determined from  the qualitative knowledge (or prediction) of the first excitations (see below).
The interpolation of $s(e)$ between $e_0$ and $e_\infty$ is thus very strongly constrained by physical considerations.
All these thermodynamic conditions do exist in a canonical description of statistical mechanics and they should equally constrain physical expansions of the free energy versus temperature and their extrapolation.  
Their implementation in the computation is never done (it would be difficult, if not impossible to do) whereas it is very simple in the present approach.

In this paper, we show that this approach  in the microcanonical ensemble allows the construction of $s(e,h)$ and as a consequence of all thermodynamic properties at all temperatures in zero and moderate magnetic fields. 
We concentrate on the spin susceptibility $\X(T)$ and test the method on two cases where the exact  function is known: the gapped one-dimensional Ising model and the gapless one-dimensional XY-model.
 Then, we address two open problems:  the antiferromagnetic Heisenberg model on the kagome lattice (supposed to be gapped\cite{Yan2011a,Depenbrock2012}) and the supposed-to-be gapless cuboc2 spin liquid phase in the $J_1$-$J_2$-$J_d$ model on the same lattice.\cite{faak2012,Bernu2013}

We consider a system of $N$ spin-$1/2$ in a constant magnetic field $B$ in the $z$-direction.
The hamiltonian reads:
\begin{align}
	\H_h=\H_0 - h S_z,
\end{align}
where $\H_0$ is a spin hamiltonian, $S_z$ the total spin along $B$, $h=mB$ and $m=g\mu_B$.
In the following, $h$ will be considered as a parameter.
The free energy per spin $f_h$ reads:
\begin{align}
\beta f_h =-\frac1N\ln \Tr\,e^{-\beta \H_h} \; \textrm{with}  \;\beta=1/T.
\end{align}
At fixed $h$, the entropy per spin $s_h$ and the energy per spin $e_h$ are given by
 \begin{align}
\label{EQ-defes}
	e_h=f_h-T\left.\frac{\partial  f_h}{\partial T}\right|_h \; \; \textrm{and} \;\;
	s_h=-\left.\frac{\partial f_h}{\partial T}\right|_h.
\end{align}
From the series expansion of $f(T,h)$ in $T$ and $h$, we first compute the HT-series of $f_h$ at fixed $h$, and from now, each function is evaluated at this $h$.
The HT series $e_h$ and $s_h$ are deduced, and
elimination of $\beta$  between them leads to the series expansion (SE) $s_h^\SE(e)$ of $s_h(e)$ around $e_\infty$ (see supplementary material).

The next step consists to extrapolate $s_h^\SE(e)$ down to $e_{h,0}$.
In the absence of a phase transition $s_h(e)$ is indeed analytic on $] \EG, 0]$,  but  singular at the boundary $\EG$, as $s_h'(e) = 1/T \to\infty$,  when  $e \to \EG$.
The  key point introduced in \cite{Bernu2001b,Misguich2005a} is then to build from $s_h(e)$ a  function $G_h(e)$ defined on $[\EG, e_\infty]$   removing  this singularity.
This can be achieved by noting that two main kinds of leading singularities are met.
If the system is gapless with a specific heat $C_v(T)_{T\rightarrow 0}\propto T^{\alpha}$, then $s_h(e\rightarrow \EG)  \propto (e - \EG) ^{\frac{\alpha}{\alpha +1}}$, and  we choose
\begin{equation}
\label{EQ:Gegapless}
	G_h(e)=	\frac{{s_h(e)}^{\frac{\alpha +1}{\alpha}}}{e-\EG}.
\end{equation}
If the system is gapped, with $C(T)\propto  \frac{1}{T^2}e^ {\Delta_h /T}$,  then  the singularity of $s_h(e \rightarrow\EG) \propto -(e-\EG)\ln(e-\EG)/\Delta_h$,
and we can choose
\begin{equation}
\label{EQ:Gegapped}
	G_h(e)=	(e-\EG)\left(\frac{s_h(e)}{e-\EG}\right)',
\end{equation}
where  the $'$ denotes the derivation  with respect to $e$.
$G_h(e)$  is a smooth function on $[\EG, e_\infty]$ easier to extrapolate.
This is done as follows:
from the series expansion of $s_h^\SE(e)$ at $e_\infty$, we deduce the series expansion of $G_h^\SE(e)$ and build the Pad\'e approximants (PA) $G_h^\PA(e)$.
The inversion of Eq.\ref{EQ:Gegapless} or \ref{EQ:Gegapped}
 gives for each PA a function $s_h^\PA(e).$\cite{SPA} By construction this method preserves both the exact information coming from the high temperature series and the supposed-to-be correct behavior at $\EG$.
{\sl At this stage any  unphysical PA, i.e. not verifying $s_h(e)>0$, $s_h'(e)>0$ and $s_h''(e)<0$ is discarded.}
The method is considered  as successful when most of the  {\sl physical} PAs coincide for $e\in[\EG,0]$.
This  criterium is a way to select the more robust approximation, and extract the more plausible information from the restricted amount of data: it is  a {\sl soft} measurement of the self-consistency of this approach (see ref.\cite{Misguich2005a} for example).
Heuristically, we noticed that spoiling the appropriate regularization at $\EG$ prevents the obtention of many coincident PAs and gives erratic results when increasing the length of the input HT series.

In order to evaluate the magnetic susceptibility $\X(T)$, we need  $f_h(T)$.
Using $s_h'(e) $= $1/T $= $\beta$, we  compute $e_h(\beta)$ from $s_h^\PA(e)$ and $f_h^\PA(\beta)=e_h(\beta)-Ts_h^\PA(e_h(\beta))$.
$\X$ is given by
\begin{align}
\label{EQ-CHI}
	\X=-\left.\frac{\partial^2 f_h}{\partial B^2}\right|_{T}=-m^2\left.\frac{\partial^2 f_h}{\partial h^2}\right|_{T},
\end{align}
where the second derivative of $f_h$ with respect to $h$ is obtained by finite differences of the same PA at different $h$.
The results presented here have been obtained  from series expansion of $f(T,h)$   at order 4 in $h$, this limits the range of applicability to small magnetic fields (not a conceptual limit just a current technical one).

\begin{figure}
\begin{center}
\hspace{-0.75cm}
\includegraphics[scale=0.55]{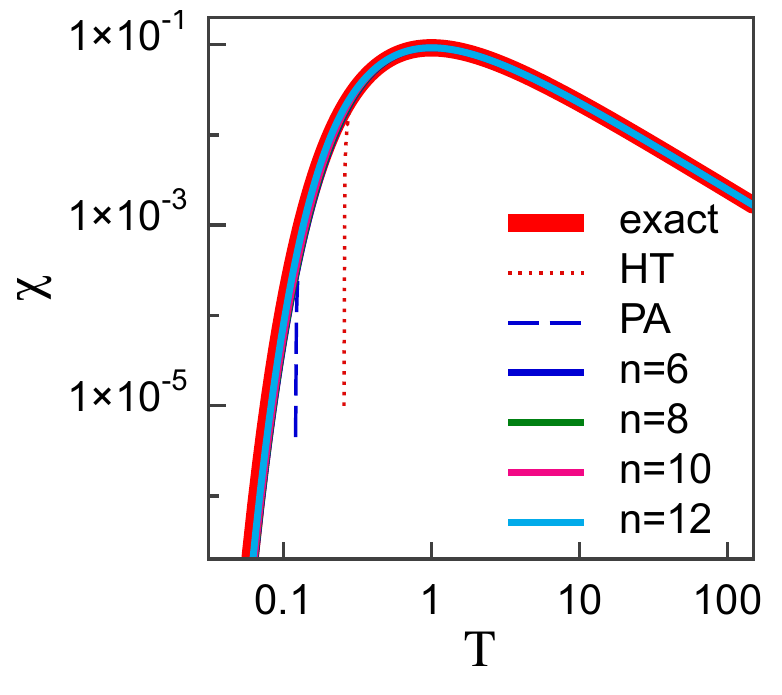}
\includegraphics[scale=0.55]{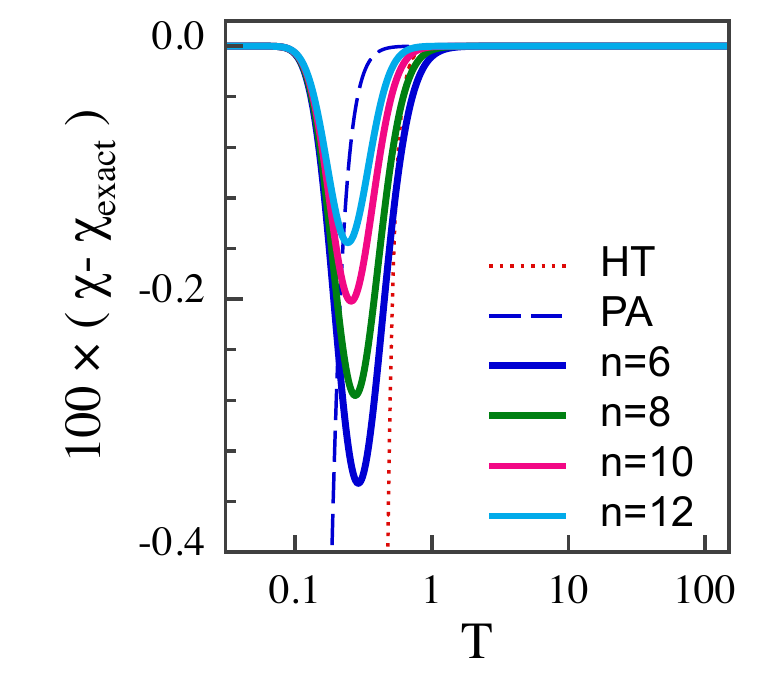}\\
\hspace{-0.75cm}
\caption[99]{(Color online) Comparison between the present method and exact results for the longitudinal  spin  susceptibility of 1D-Ising model.
Left: our results compared to the exact solution.
Right: differences between exact $\X$ and the various approximations.
HT (dotted line) stands for the HTE at order 12,
PA  (dashed line) stands for the [6-6]-Pad\'e approximant,
the other curves stand for the present method using HTE at various order $n$.
}
\label{FIG-Ising1D-Xs}
\end{center}
\end{figure}

\textit{Gapped systems: the longitudinal  spin susceptibility of the 1D-Ising model at $h=0$.}\\
The Hamiltonian is $\H_0=\sum_i 2 S_{i,z}S_{i+1,z}$.
We use  the regularizing function defined in Eq.\ref{EQ:Gegapped}.
With the exact value $e_0(h)=  -1/2$, and an HTE at order 4 only, $\X$ is already reproduced within 1\%.
Increasing the order  of the series decreases  both the maximum error and the range of temperature where the errors are non negligible.

As in most cases $e_0(h)$ is unknown, we  have also considered it as a free parameter to check the method.
$e_0$ is then adjusted by maximizing the number of $s_h^{PA}(e)$.
This criterium is very accurate and gives $e_0\in [-1/2 - 10^{-9} , -1/2+ 10^{-7}]$.
The error  on $\X$ for the 12-order  HT-series is less than $2\times10^{-3}$ (see Fig.\ref{FIG-Ising1D-Xs}),
and we find a gap to be the exact value 1, within an error of $10^{-9}$.

\begin{figure}
\begin{center}
\hspace{-0.75cm}
\includegraphics[scale=0.55]{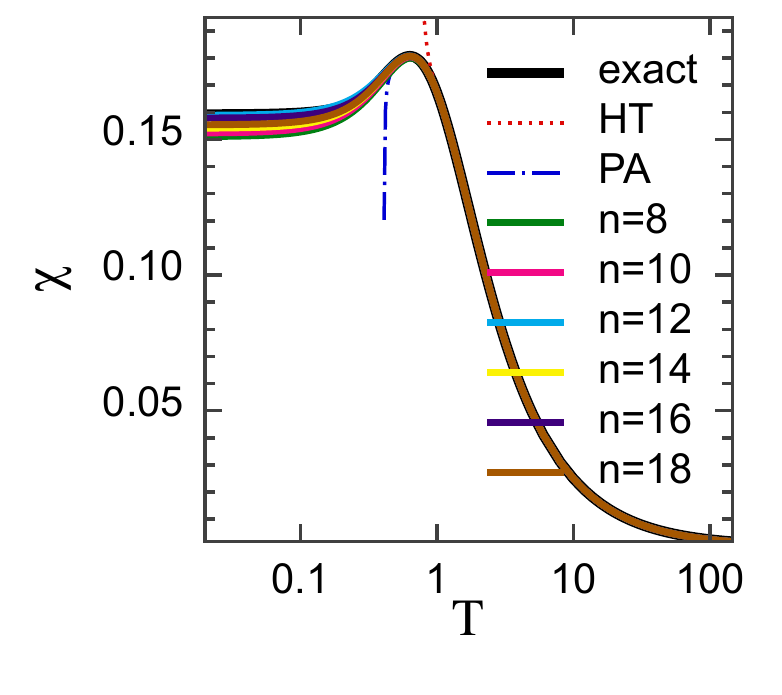}
\includegraphics[scale=0.55]{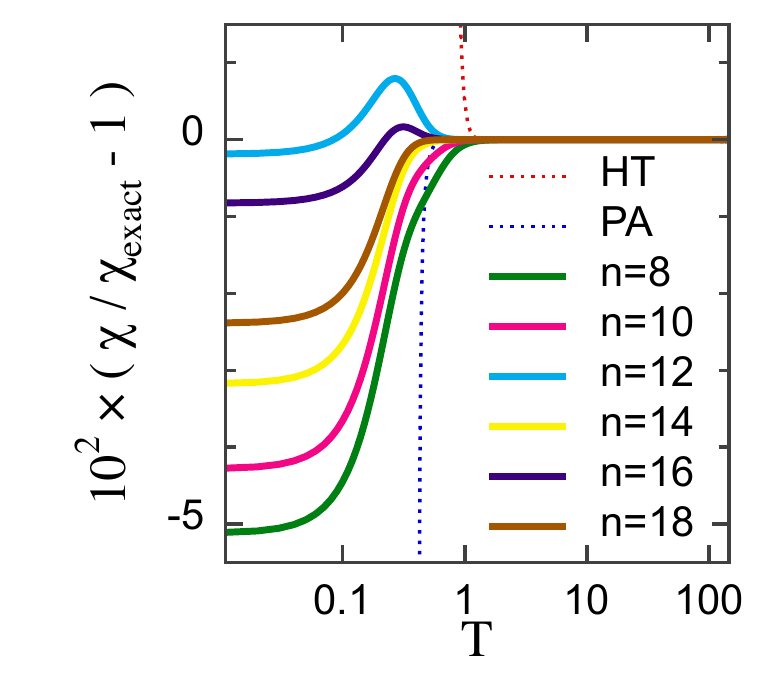}
\caption[99]{(Color online) Same as Fig.\ref{FIG-Ising1D-Xs} for the transverse spin susceptibility of 1D-XY model, where $e_0(h)$ is left free to adjust.}
\label{Fig:XYe0free}
\end{center}
\end{figure}

\textit{Gapless model: the transverse spin susceptibility of the 1d XY-model.}\\
The Hamiltonian reads $\H_0=\sum_i 2 (S_{i,x}S_{i+1,x}+S_{i,y}S_{i+1,y})$.
Exact results have been obtained by Katsura.\cite{Katsura1962}
The specific heat is linear in $T$ at low temperature, thus
the singularity of $s_h(e)$ around $\EG$ is regularized through Eq.\ref{EQ:Gegapless}.
Using the exact value of $\EG$ leads to the exact value of $\X(T=0)$ and values of $\X$ within an error of less than $1\%$ in the whole range of temperature for a 12-term {\b HT-}series.
Leaving $\EG$ as a free parameter, the errors do not exceed a few percents in the whole range of temperature (see Fig.\ref{Fig:XYe0free}).
\begin{figure}
\begin{center}
\includegraphics[width=0.35\textwidth]{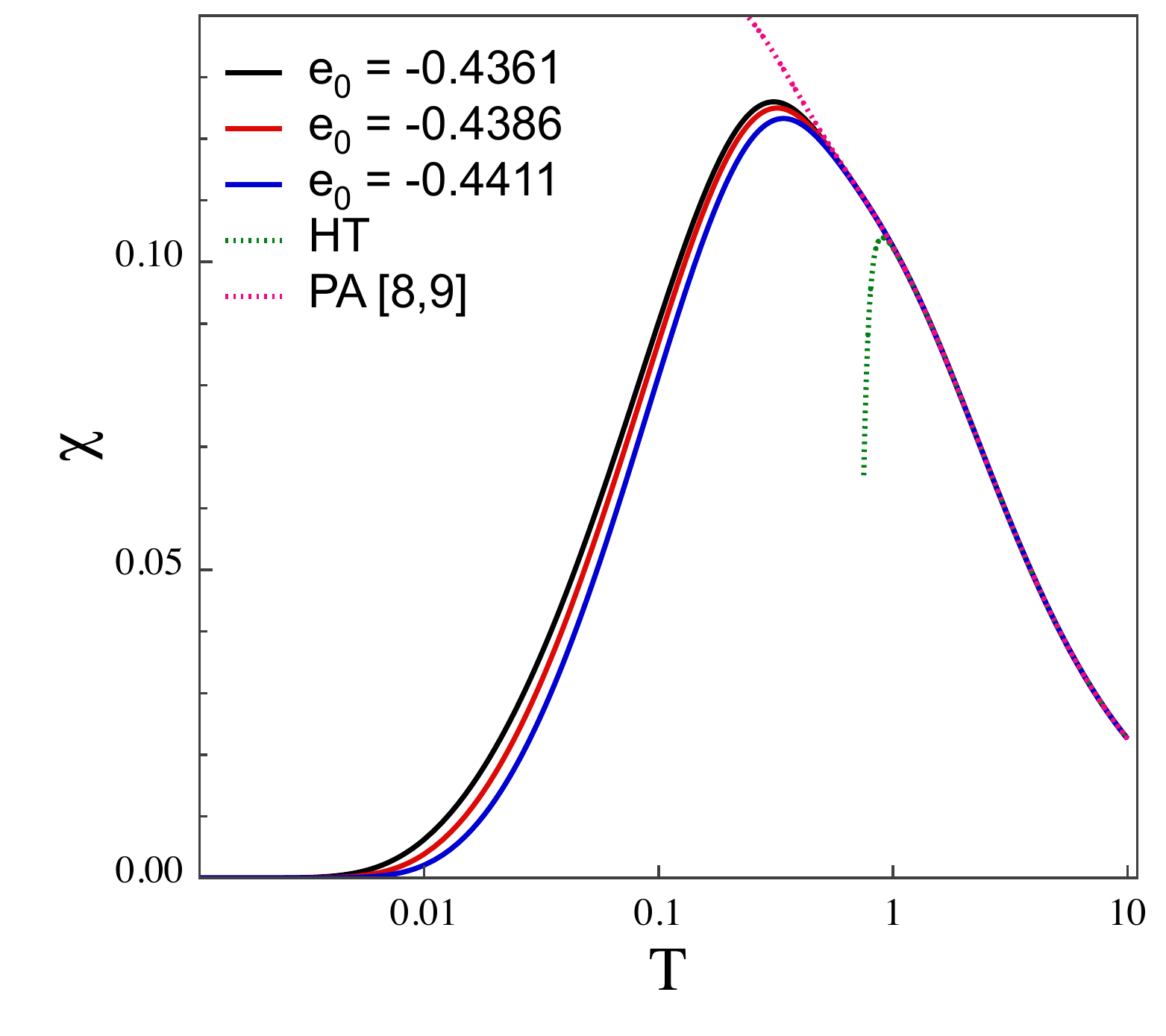}
\includegraphics[width=0.35\textwidth]{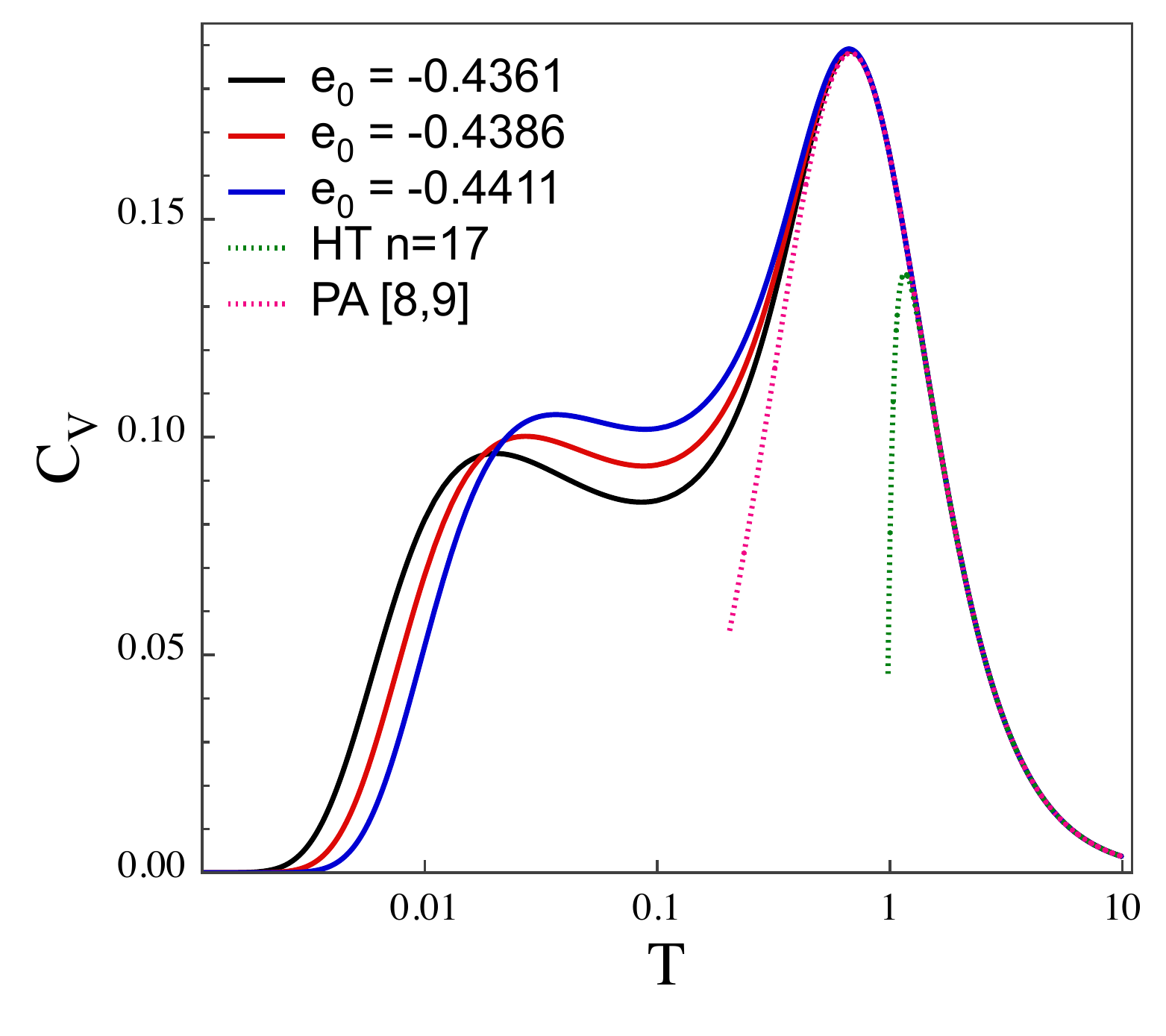}
\caption[99]{(Color online) Spin susceptibility  $\chi$ and specific heat $C_v$ of the antiferromagnetic Heisenberg model on the kagome lattice.  Shown in the figures are the HT series expansion to order 17 (green dotted lines), the best Pad\'e approximant of this simple series (magenta dotted line), and  the results of the present interpolation (full lines). The sensitivity of the interpolation to the ground-state energy  $e_0$ is displayed on both quantities. The full red curves are associated  to the best commonly admitted value of $e_0$ (see text).
\label{FIG:kag}
}
\end{center}
\end{figure}

\textit{Antiferromagnetic Heisenberg model on the kagome lattice}.\\
The spin-1/2 antiferromagnetic Heisenberg model on the kagome lattice,
$\H_0=\sum_{<i,j> }\textbf{S}_{i}.\textbf{S}_{j}$,
is a quintessential example of  the effects of both geometric frustration and quantum fluctuations  pushed to their limit.  After many studies, decisive progresses in 2D DMRG  have ascertain the value of the ground-state energy of this system, $e_0= - 0.4386(5)$, and estimate the spin gap of the order of 0.13(1).\cite{Yan2011a,Depenbrock2012}
Early HTE of Eltsner et Young \cite{Elstner1994}, extended in this work to order 17 give a first idea of the HT behavior of the thermodynamical quantities.
The HTE diverges around $T=1$ and the Pad\'e approximants of this series diverge below $T=0.4$ (see Fig.\ref{FIG:kag}).

With the hypothesis of a gapped system, $G(e,h)$ is built using Eq.(\ref{EQ:Gegapped}).
Results displayed in  Fig.\ref{FIG:kag} are obtained by fixing $e_0$ to its best known value ($e_0= - 0.4386$) (red curve) and to two extreme values which differ by 5 standard deviations from the the present best DMRG estimate.
For a given value of $e_0$, the differences between the various $\X^\PA$ are less than the thickness of the lines.
For this range of ground state energies, we find a gap 0.03(1), significantly smaller than the gap obtained in the DMRG approach.
Compared to exact diagonalisations (ED) on 18 and 24-spin samples\cite{Elstner1994,Misguich2007}, we find, at the thermodynamic limit, a smaller value for the maximum of $\X$ ($\sim 0.12$).
In ED, the spin-spin correlations are indeed overemphasized by the very small lengths of the samples.
The existence of a low temperature shoulder in the specific heat is confirmed.
Unfortunately, comparison with experimental date of Herbertsmithite cannot be done because of sizable Dzyaloshinskii-Moriya interactions that change the low energy spectrum of excitations and probably close the gap.\cite{Zorko2008,Cepas2008}

{\textit{Spin susceptibility of Kapellasite}}.\\
\begin{figure}
\begin{center}
\includegraphics[width=0.35\textwidth]{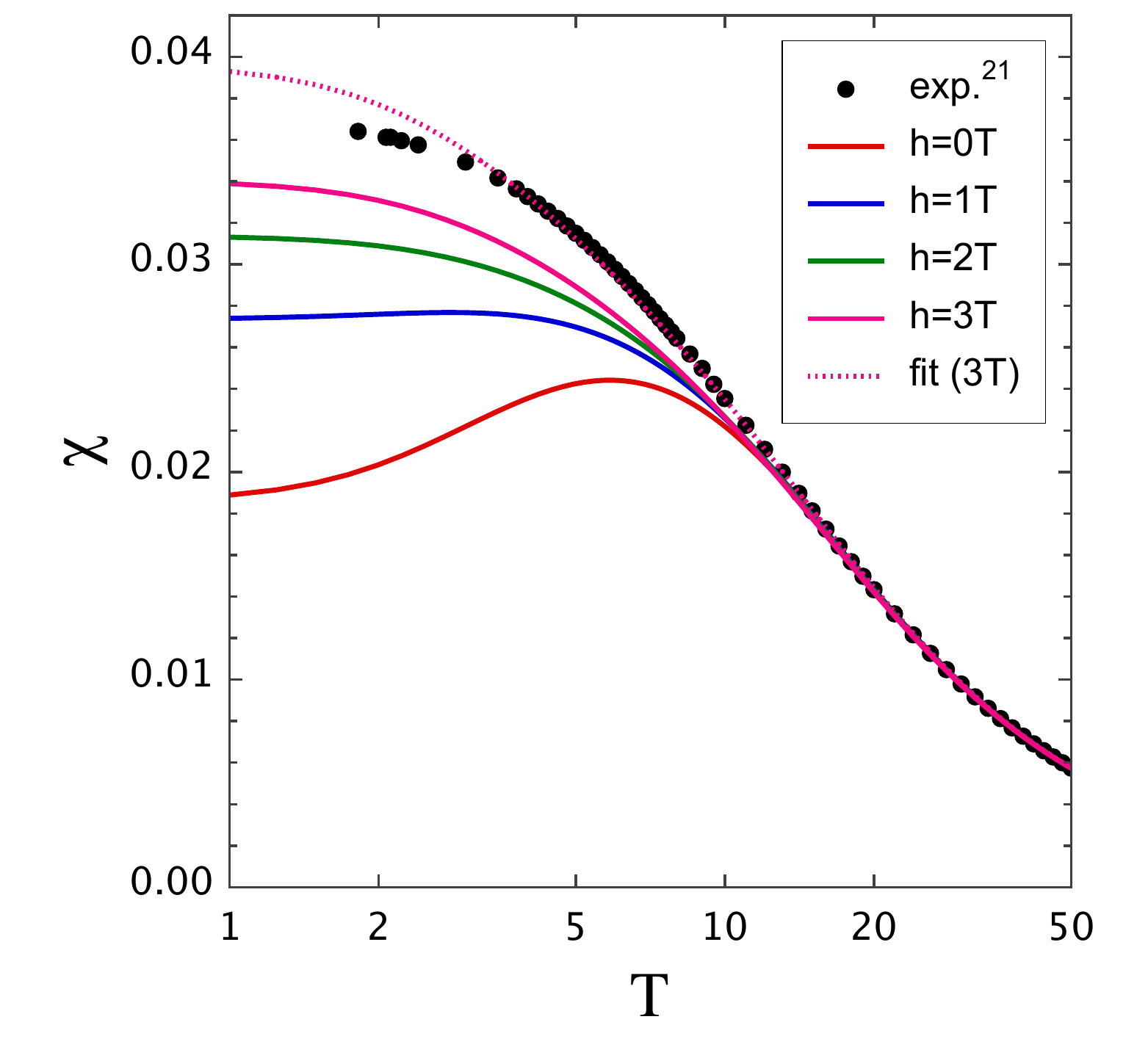}
\caption{
(Color online) Spin susceptibility comparison between experiment\cite{EDWIN} at 5T and the $J_1$-$J_2$-$J_d$ model for different values of the magnetic field 0, 1, 2, 3T with $J_i (K)=[-12, -4,15.6]$ and the new fit (see text).
\label{Fig:kapellasite}
}
\end{center}
\end{figure}
This material is in a Mott phase and its properties are analyzed using a spin-1/2 Hamiltonian on the kagome lattice:\cite{faak2012,Bernu2013}
\begin{equation}
	\H_0= J_1 \sum_{\langle i,j\rangle }\textbf{S}_{i}.\textbf{S}_{j} + J_2\!\!\sum_{\langle\langle i,j\rangle\rangle }\!\! \textbf{S}_{i}.\textbf{S}_{j}  + J_d \!\!\!\!\sum_{\langle\langle\langle i,j\rangle\rangle\rangle }\!\!\!\!\textbf{S}_{i}.\textbf{S}_{j}
\end{equation}
where $J_d$ is the third-neighbor exchange energy across the hexagon.
The best set of parameters, obtained from a fit of the spin susceptibility down to $T= 17.5K$,
reads $J_1=-12K$  , $J_2=-4K$, and $ J_d=15.6K$.\cite{Bernu2013}
The low-temperature specific heat is experimentally known to be  $\propto T^2$, we then use Eq.\ref{EQ:Gegapless} to regularize $s(e)$ and adjust $\EG$.
The full curves of Fig.\ref{Fig:kapellasite} are obtained for the above-mentioned best set. As expected they indeed agree with experiments down to 17.5K.
We see an increasing disagreement with experimental data when going to lower temperatures.
Part of the disagreement can be  assigned to the magnetic field, which has a large effect in this system with competing interactions (Fig.\ref{Fig:kapellasite}).
Up to now, the HTE for this model are available  to order 4 in $h$ only, which limits the evaluation of $\X(h)$ to $h/|J| \lesssim 0.25$, that is a magnetic field less than 3 Tesla, while
experimental data are at 5 Tesla.
Nevertheless, with a small change of the coupling constants $J_1= -12$\,K  $J_2 = -5.2$\,K and $J_d = 16.4$\,K, and 3 Tesla magnetic field, we can fit almost all experimental data.
A small disagreement persists at the lowest temperature, where the magnetic field effects are the most important.
The uncertainties on the set of parameters is considerably reduced with this present method because we use experimental data at all temperatures and  include the effect of the magnetic field.
For Kapellasite, experimental data at lower field and/or longer series in $h$ will help to achieve a better determination of the parameters.

In this paper we have proposed a method to extend the HTE of the spin susceptibility down to  $T=0$, based on a reconstruction of the entropy versus energy per spin. We have checked the method against gapless and gapped exact models: the largest deviations from the exact results are of the order of $10^{-2}$ or better with an original HT series expansion of ten terms.  We have applied this method to open problems on the kagome lattice.
Being not limited by finite size effects, we believe in the accuracy of the present method compared to that of exact diagonalisations, specially at low temperature. We have equally shown that this approach can be used to compute the spin  susceptibility in a finite magnetic field, which is a major point for the comparison between models and experimental squid data.\cite{WEB}
The method is general in its principle and can be applied straightforwardly to other models, whatever the size of the spin, as much as the high temperature expansion of the free energy per spin is available.\cite{lohmann2014}
This opens a large range of interesting studies such as Herbertsmithite confronted to Heisenberg model with Dzaloshinskyi-Moryia interactions, Volborthite with spatially anisotropic couplings, spin ices ...
\cite{Shimizu2003,mendels2007,helton07,Yamashita2008,Yamashita2011,faak2012,Han2012,Han2012a}
A further conceptual question has not been studied in the present work: is this approach able to deal with critical phase transitions? This might be possible for temperatures larger than $T_c$ in as much as the correct diverging behavior at $T_c$ is taking into account, but this is probably more delicate than the present work as these divergences are singularities in the derivatives of $s(e)$ (see supplementary material). Building of a suitable regularization function and benchmarking the method is a new subject in itself, beyond the scope of this Letter.

We thank G. Misguich for fruitful discussions.
We acknowledge the support of the French ministry of research through the ANR grant   "SpinLiq".


%

\newpage
\section*{Supplementary Material}
In this supplementary material we give some details on how the series $s(e)$ is obtained.

A section is devoted to the 3D ferromagnetic Heisenberg model on the bcc lattice where we show that even if the present method has not been adapted for a critical phase transition, it nevertheless give some evidence of a critical temperature within 10\% of the correct result.

HT-series are given in the last section.

\section{Evaluation of $s(e)$ at high temperature}
We consider $N$ spins on a given lattice in an external constant magnetic field $B$.
The Hamiltonian reads:
\begin{align}
\label{EQ-def-H}
	\H=&\H_0-h \sum_iS_{z,i}
\end{align}
where $\H_0$ is a spin hamiltonian depending on a few exchange parameters, $h=m B$ and $m=g\mu_B$.
The partition function is $\Z=\Tr e^{-\beta\H}$, with $\beta=1/T$ and the free energy, the energy and the entropy per spin are obtained from
\begin{align}
\label{EQ-def-fh}
	\beta f_h=&-\frac1N\ln \Z_h\\
	e_h=&f_h-T\left.\frac{\partial  f_h}{\partial T}\right|_h=-\left.\frac{\partial \ln Z}{\partial \beta}\right|_h \\
	s_h=&-\left.\frac{\partial f_h}{\partial T}\right|_h=-\beta^2\left.\frac{\partial \frac1\beta\ln Z}{\partial \beta}\right|_h.
\end{align}

The HT-series of $\ln \Z_h$ as a function of $\beta$ and $h$ is written as
\begin{align}
\label{EQ-deflnZ}
	\frac1N\ln \Z_h^\HT	&=L_0+L_2\frac{(\beta h)^2}{2^22!}+L_{4}\frac{(\beta h)^4}{2^44!}+...\\
\label{EQ-defL2nu}
	L_{2\nu}&=\sum_{i=0}^n L_{2\nu,i}\frac{\beta^i}{i!}
\end{align}
where $L_{2\nu,i}$ are homogeneous polynomials of the exchange parameters.
In the following, $h$ is a parameter and $\ln \Z_h^\HT$ is evaluated numerically as a polynomial in $\beta$, as well as all other functions. Thus we have
\begin{align}
	\ln Z_h^\HT&=\ln2+\sum_{i=2}^n L_i \beta^i\\
\label{EQ-def-ebeta-HT}
	e_h^\HT(\beta)&=-\sum_{i=1}^{n-1}(i+1)L_{i+1}\beta^i\\
\label{EQ-def-sbeta-HT}
	s_h^\HT(\beta)&=\ln2-\sum_{i=2}^{n}(i-1)L_{i}\beta^i
\end{align}
where $n$ is the order of HT-series and
$L_1=-e_\infty=0$ as we can define $\H_0$ such as the mean energy at infinite temperature is 0.

The series $\se^\SE(e)$ is obtained by eliminating $\beta$ between $e_h^\HT$ and $s_h^\HT$:
\begin{align}
\label{EQ-def-se-HT}
	\se_h^\SE(e)&=\ln2+\sum_{i=2}^{n}\se_i e^i,\qquad \se_i=-\frac{b_i}{2^iL_2^{2i-3}},
\end{align}
we get for the first term
$b_2=1$,
$b_3 = L_3$ and
{\small
\begin{align}
	b_4 &= \frac94L_3^2-L_4L_2,\\
	b_5 &= \frac{27}4L_3^3-6L_4L_3L_2+L_5L_2^2,\\
	b_6 &= \frac{189}8L_3^4\!-\!\frac{63}2L_4L_3^2L_2\!+\!\left[\frac{15}2L_5L_3\!+\!4L_4^2\right]L_2^2
		\!-\!L_6L_2^3.
\end{align}
}
An algorithm to get the general term reads:
\begin{enumerate}
	\item store the polynomial $e_h^\HT$ whose coefficients are $e^\HT_{h,i}=-(i+1)L_{i+1}$ for $i=0...n-1$ (Eq.\ref{EQ-def-ebeta-HT})
	\item compute and store the polynomials $\left(e_h^\HT\right)^{k}$ up to order $n$ ($\left(e_h^\HT\right)^{k}$ starts at order $k$).
	\item for $i$ from 2 to $n$ compute $\se_i$ by solving the triangular set of linear equations:
\begin{align}
		\se_i&=-\left(\frac{-1}{2L_2}\right)^i\left((i-1)L_i+\sum_{k=2}^{i-1} \se_k\,\left[\left(e_h^\HT\right)^{k}\right]_i \right)
\end{align}
where $\left[\left(e_h^\HT\right)^{k}\right]_i$ is the coefficient of order $\beta^i$ in $\left(e_h^\HT\right)^{k}$.
\end{enumerate}

\section{Evidence of a critical behavior from a $\se(e)$ analysis}
This part is an unpublished work done more than ten years ago with G. Misguich.

\begin{figure*}
\begin{center}
\includegraphics[width=0.4\textwidth]{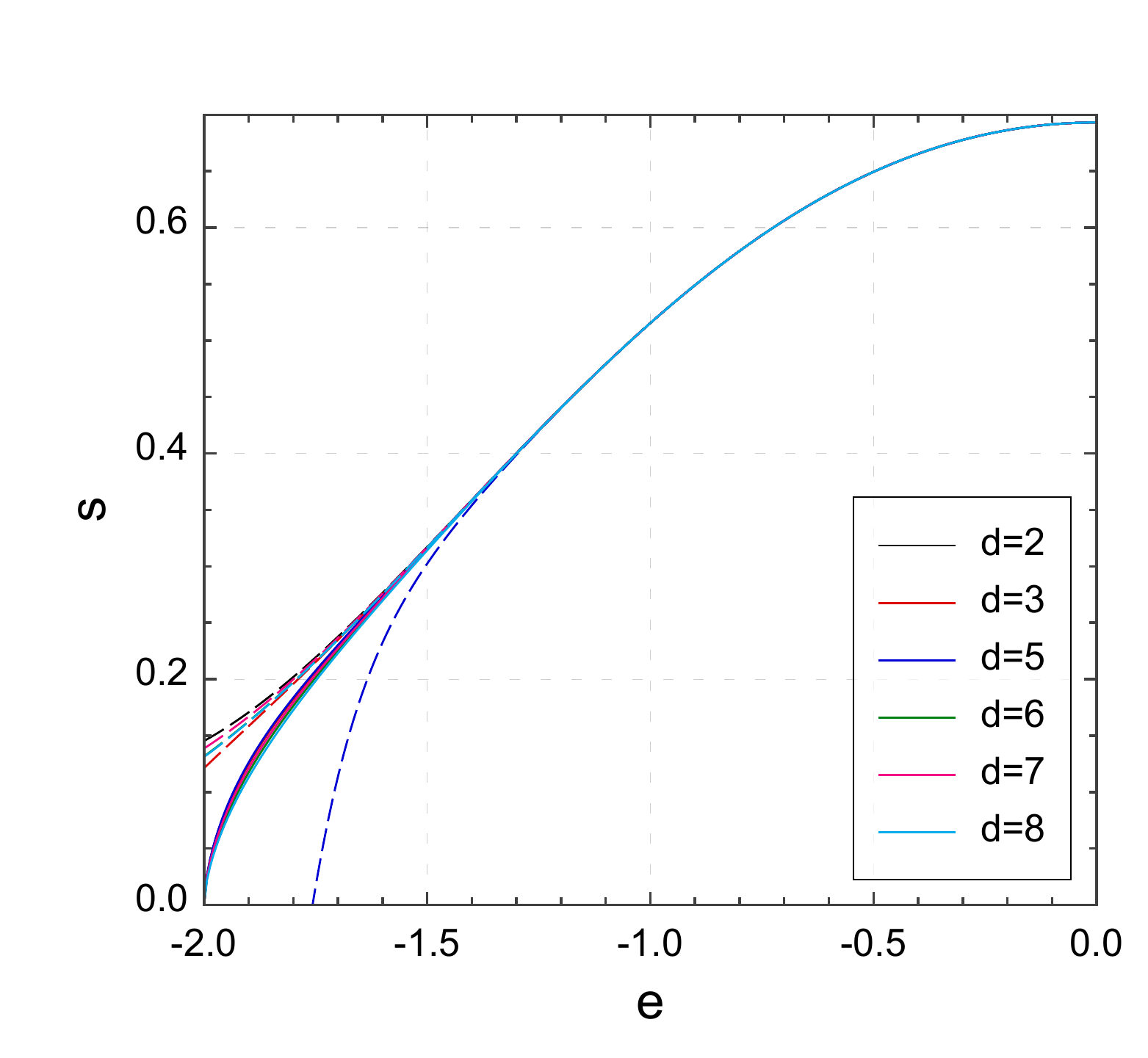}
\includegraphics[width=0.4\textwidth]{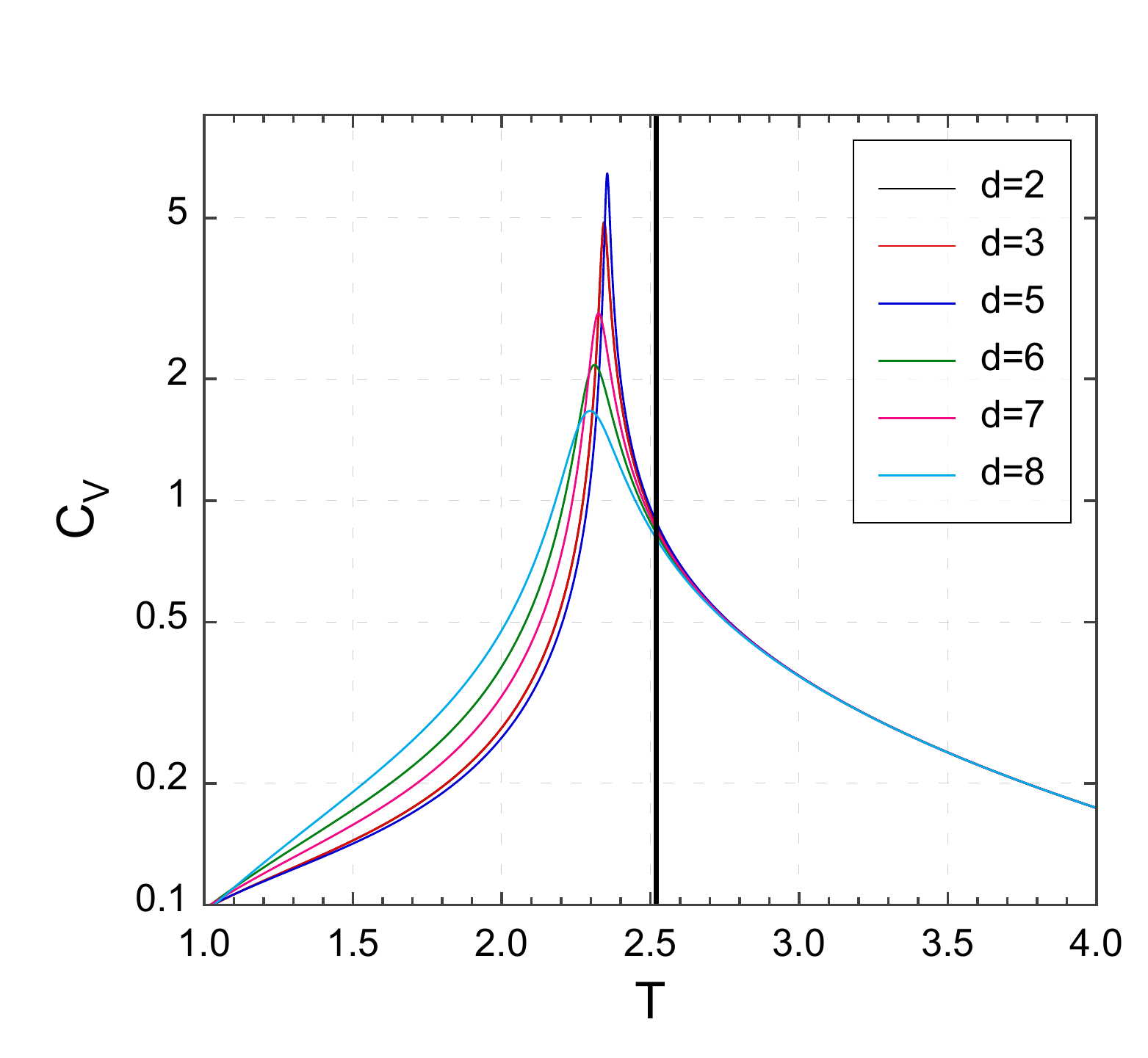}
\caption{
Thermodynamic functions for the Heisenberg model on the bcc lattice. \\
Left: $\se(e)$ the entropy per spin versus energy per spin.
Dashed lines are pade approximants $[n-d,d]$ of the series $\se^\HT(e)$ at order 14.
Full lines are the physical pade approximants, obtained by the present approach, starting at $e_0=-2$ with $\se^\PA(e_0)=0$ and ending at $e=0$ with $\se(0)=\ln2$.\\
Right: $C_V(T)$ specific heat, in log-scale, versus temperature.
$C_V(T)$ is obtained from $\se^\PA(e)$ and Eqs. (\ref{EQ-def-beta}),(\ref{EQ-def-CV-from-se}).
The vertical line shows the estimate of the critical temperature by J. Oitmaa and E. Bornilla, Phys. Rev. B p14228 (1996).
\label{BCC-FERRO}
}
\end{center}
\end{figure*}

In this section, we show that the analysis of the function $\se(e)$ of a 3D ferromagnetic spin model provides some evidence of a critical behavior.
We recall that the first derivative of $\se(e)$ is positive
\begin{align}
\label{EQ-def-beta}
	\se'(e)=\beta,
\end{align}
its second derivative is negative, and the specific heat is given by (see ref.14 of main article)
\begin{align}
\label{EQ-def-CV-from-se}
	\CV=-\frac{\se'(e)^2}{\se''(e)}
\end{align}
Suppose that $\se(e)$ is almost linear on an interval $[e_1,e_2]$.
Then $\se'(e)$ is almost constant, and $\se''(e)$ is very small and may vanish.
This results in a sharp peak in $C_V$, and a divergence if $s''(e)$ vanishes.
This is exactly what we find in the case of the three dimensional Heisenberg model on the bcc lattice ($\mathcal H=\sum_{\langle i,j\rangle} 2S_i\cdot S_j$).
In fig.\ref{BCC-FERRO}-left, the dashed lines represent the direct Pade approximants
of $\se^\SE(e)$  at order $n=14$.
We see that in the interval $[-1.3,-1]$ the function is rather flat.
Building the specific heat from those Pades approximants and Eq.\ref{EQ-def-CV-from-se}, leads to a divergence of $C_V$ around $T\approx8.8$, a temperature much larger than $T_c=2.52$ [J. Oitmaa and E. Bornilla, Phys. Rev. B p14228 (1996)].

Applying the method described in the main article with a low temperature behavior $C_V\propto T^{3/2}$ and with the knowledge of the exact ground state energy $e_0=-2$, we build the smooth Pade functions $\se^\PA(e)$ (full lines of fig.\ref{BCC-FERRO}-left).
Fig.\ref{BCC-FERRO}-right shows the corresponding specific heat, in log-scale, where we see sharp peaks at temperatures about 10\% lower than $T_c$.

Here we have a good example of the power of working with the function $\se(e)$. The constrains imposed on $\se(e)$ are so strong that even if we miss some feature (here a singularity of $\se'''(e)$ at some critical value $e_c$), we get a clear signal of its presence. In that sense, a careful analysis of the function $\se(e)$ may be sufficient to predict new physics.

Improving the present method to account properly for the singularity at $T_c$ is an interesting prospect.

\section{HT-series}
In this section we give the coefficients of the polynomials for various models.
The hamiltonians are of the form
\begin{align}
	\H_0=2\sum_{i,j} J_{ij} S_i\cdot S_j
\end{align}

\subsection{One-dimensional Ising model}
The HT-series in $\beta$ and $h$ are easily obtained from the exact partition function:
\begin{align}
	\ln Z(\beta,h)=e^{-\beta/2}\left[\cosh\left(\frac{\beta h}2\right)+\sqrt{\sinh^2\left(\frac{\beta h}2\right)+e^{2\beta}}\right]
\end{align}

\subsection{One-dimensional XY model in transverse field}
The HT-series have been deduced from the exact expression given by S. Katsura, Phys. Rev. 127 p.1508 (1962).
{\footnotesize
\begin{align}
\nonumber
	L_{0,i}=&\ln 2, 0, \frac{1}{2}, 0, -\frac{3}{4}, 0, 5, 0, -\frac{595}{8}, 0, 1953, 0, -\frac{159621}{2}, 0,
\nonumber
		 4685538, 0, -\frac{5981776515}{16}, 0, 38923847105, 0\\
\nonumber
	L_{2,i}=& 1, 0, -1, 0, 6, 0, -85, 0, 2170, 0, -87066, 0, 5045964, 0,
\nonumber
	-398785101, 0, 41213485170, 0, -5395132424110, 0\\
\nonumber
	L_{4,i}=&-2, 0, 8, 0, -102, 0, 2480, 0, -96740, 0, 5504688, 0,
\nonumber
	 -429460878, 0, 43961050848, 0, -5712493154940, 0,
\nonumber
	  918357221004080, 0
\end{align}
}

\subsection{One-dimensional Ising model in transverse field}
Same as previous subsection.
{\footnotesize
\begin{align}
\nonumber
	L_{0,i}=&\ln 2, 0, 1, 0, -2, 0, 16, 0, -272, 0, 7936, 0, -353792,
\nonumber
			0, 22368256, 0, -1903757312, 0, 209865342976, 0\\
\nonumber
	L_{2,i}=& 1, 0, -\frac43, 0, \frac{48}5, 0, -\frac{1088}7, 0, \frac{39680}9, 0, -\frac{2122752}{11}, 0,
\nonumber
		 \frac{156577792}{13}, 0, -\frac{15230058496}{15}, 0, \frac{1888788086784}{17}, 0,
\nonumber
  -\frac{290888851128320}{19}, 0\\
\nonumber
	L_{4,i}=&-2, 0, \frac{48}5, 0, -\frac{4896}{35}, 0, \frac{79360}{21}, 0, -\frac{1768960}{11}, 0, \frac{1409200128}{143},
\nonumber
	 0, -\frac{53305204736}{65}, 0, \frac{7555152347136}{85}, 0, -\frac{3926999490232320}{323}, 0,
  	\\&
\nonumber
	 \frac{272332392921825280}{133}, 0
\end{align}
}
\subsection{Kagome lattice}
We give here the series for various models on the kagome lattice.\\
\subsubsection{$J_1$ model}
{\footnotesize
\begin{align}
\nonumber
	L_{0,i}=&\ln 2, 0, \frac{3}{2}, 0, -\frac{51}{4}, 0, \frac{2139}{4}, -\frac{483}{2}, -\frac{408459}{8}, 77787, \frac{34646571}{4},
\nonumber
		 	 -28988091, -\frac{18289265325}{8}, \frac{27736662603}{2}, \frac{6898225312557}{8},
	  	\\&
\nonumber
			-\frac{34038179733525}{4}, -\frac{7003408736492787}{8} , 6576601728421692\\
\nonumber
	L_{2,i}=& 1, -2, 4, -3, -4, -202, 1513, 13844, -186286, -2329677,
\nonumber
	 44494564, 568071766, -15809083661, -\frac{386791997479}{2},
  	\\&
\nonumber
	  7857174705265, 84970643937857, -5176017551551186
\end{align}
}
\subsubsection{$J_1$-$J_2$ model}
First and second neighbor model on the kagome lattice.
The polynomials homogeneous in the coupling parameters are defined as
\begin{align}
\label{EQ-defJ1J2}
	L_{2\nu,i} =\sum_{j=0}^i L_{2\nu,i,j} J_1^{i-j} J_2^j
\end{align}
\underline{\bf$ L_{0,ij}$}
{\footnotesize
\begin{align}
\nonumber
L_{0,0,j}=&  \ln2 \\
\nonumber
L_{0,1,j}=&  0, 0 \\
\nonumber
L_{0,2,j}=&  \frac{3}{2}, 0, \frac{3}{2} \\
\nonumber
L_{0,3,j}=&  0, -\frac{9}{2}, 0, 0 \\
\nonumber
L_{0,4,j}=&  -\frac{51}{4}, 12, 0, 0, -\frac{51}{4} \\
\nonumber
L_{0,5,j}=&  0, 165, -\frac{285}{2}, \frac{165}{2}, 0, 0 \\
\nonumber
L_{0,6,j}=&  \frac{2139}{4}, -819, -\frac{1449}{2}, 663, 315, 0, \frac{2139}{4} \\
\nonumber
L_{0,7,j}=&  -\frac{483}{2}, -\frac{30681}{2}, 25053, -10962, \frac{5691}{2}, -\frac{44289}{4}, 0, -\frac{483}{2} \\
\nonumber
L_{0,8,j}=&  -\frac{408459}{8}, 110334, 168378, -362052, \frac{329469}{2}, -27396,
\nonumber
	-11184, 0, -\frac{408459}{8} \\
\nonumber
L_{0,9,j}=&  77787, 2561598, -\frac{12045861}{2}, 2492136, 1329210, 1721169,
\nonumber
	-974223, \frac{4060989}{2}, 0, 77787 \\
\nonumber
L_{0,10,j}=&  \frac{34646571}{4}, -25289685, -50748525, 166662870, -\frac{253226685}{2},
\nonumber
	39027270, -\frac{107801715}{2}, 14131635, -2125020, 0, \frac{34646571}{4} 
\end{align}
}
\underline{\bf$ L_{2,ij}$}
{\footnotesize
\begin{align}
\nonumber
L_{2,0,j}=&   1\\ 
\nonumber
L_{2,1,j}=&  -2, -2\\ 
\nonumber
L_{2,2,j}=&  4, 16, 4\\ 
\nonumber
L_{2,3,j}=&  -3, -81, -96, -3\\ 
\nonumber
L_{2,4,j}=&  -4, 172, 1094, 432, -4\\ 
\nonumber
L_{2,5,j}=&  -202, 605, -5595, -11645, -1360, -202\\ 
\nonumber
L_{2,6,j}=&  1513, 4104, -14253, 132562, 98088, 6768, 1513\\ 
\nonumber
L_{2,7,j}=&  13844, -151620, 139083, -\frac{189371}{2}, -\frac{4759391}{2}, -\frac{1275309}{2},
\nonumber
	-106036, 13844\\ 
\nonumber
L_{2,8,j}=&  -186286, -137536, 6324260, -10846320, 16142374,
	32058256, 4069692, 764352, -186286\\ 
\nonumber
L_{2,9,j}=&  -2329677, 26960814, -56758545, -115342752, 319743243,
	-580389138, -314815344, -55709883, 7338168, -2329677\\ 
\nonumber
L_{2,10,j}=&  44494564, 32699900, -1607336300, 4682885400,
	-1969984450, -3722864284, 12819641560, 2641862210,
	853326455, -109501560,
\\ \nonumber
	&44494564 
\end{align}
}
\underline{\bf$ L_{4,ij}$}
{\footnotesize
\begin{align}
\nonumber
L_{4,0,j}=&   -2 \\
\nonumber
L_{4,1,j}=&    16, 16 \\
\nonumber
L_{4,2,j}=&    -116, -320, -116 \\
\nonumber
L_{4,3,j}=&    744, 4284, 4512, 744 \\
\nonumber
L_{4,4,j}=&    -4312, -44336, -101080, -52416, -4312 \\
\nonumber
L_{4,5,j}=&    25616, 370880, 1606560, 1893640, 529280, 25616 \\
\nonumber
L_{4,6,j}=&    -195248, -2709360, -19245948, -45281720, -30254952,
	-4898880, -195248 \\
\nonumber
L_{4,7,j}=&    1664756, 22589784, 180910170, 774556076, 1076124056,
	428284248, 46001648, 1664756 \\
\nonumber
L_{4,8,j}=&    -8817028, -254148640, -1625660416, -9686835696,
	-26227733120, -22350160640, -5601590400, -482074368,
	-8817028 \\
\nonumber
L_{4,9,j}=&    -3185856, 2368385640, 20923820496, 97265440680,
	446985312264, 773024430096, 415486343928, 72171633312, \\
\nonumber
	&5155399872, -3185856 \\
\nonumber
L_{4,10,j}=&    -732385616, -4871423920, -293070042320,
	 -1254470305800, -5403290969680, -18193823967520,
	 -20183796938300,
	 &\\\nonumber	&
	 -7123711581640, -970730953660,-44469171360, -732385616
\end{align}
}

\subsubsection{$J_1$-$J_3$ model}
First and third neighbor (along a line) model on the kagome lattice (see Eq. \ref{EQ-defJ1J2}).\\
\underline{\bf$ L_{0,ij}$}
{\footnotesize
\begin{align}
\nonumber
L_{0,0,j}=&   \ln2 \\
\nonumber
L_{0,1,j}=&  0,0 \\
\nonumber
L_{0,2,j}=&  \frac{3}{2}, 0, \frac{3}{2} \\
\nonumber
L_{0,3,j}=&  0, -\frac{9}{2}, 0, \frac{3}{2} \\
\nonumber
L_{0,4,j}=&  -\frac{51}{4}, 12, -\frac{27}{2}, 0, -\frac{21}{4} \\
\nonumber
L_{0,5,j}=&  0, \frac{375}{2}, -120, -\frac{45}{2}, 0, -\frac{75}{2} \\
\nonumber
L_{0,6,j}=&  \frac{2139}{4}, -864, 666, 303, \frac{5319}{4}, 0, \frac{195}{4} \\
\nonumber
L_{0,7,j}=&  -\frac{483}{2}, -\frac{38997}{2}, \frac{44037}{2}, -\frac{50757}{4}, -2394, 5208, 0, \frac{7791}{4} \\
\nonumber
L_{0,8,j}=&  -\frac{408459}{8}, 122946, -\frac{44715}{2}, -137034, -\frac{1021305}{4}, 63432,
	-\frac{199629}{2}, 0, \frac{28779}{8} \\
\nonumber
L_{0,9,j}=&  77787, \frac{7143309}{2}, -5858379, \frac{12197421}{2}, 372114, \frac{637119}{2},
	-132957, -1391877, 0, -\frac{337011}{2} \\
\nonumber
L_{0,10,j}=&   \frac{34646571}{4}, -29580030, -\frac{66170655}{4}, 73224780, \frac{52677345}{2},
	-43400340, 82371840, -2740830, \frac{37148955}{4}, 0, -\frac{4812639}{4} 
\end{align}
}
\underline{\bf$ L_{2,ij}$}
{\footnotesize
\begin{align}
\nonumber
L_{2,0,j}=&  1 \\
\nonumber
L_{2,1,j}=&  -2, -2 \\
\nonumber
L_{2,2,j}=&  4, 16, 4 \\
\nonumber
L_{2,3,j}=&  -3, -81, -96, -8 \\
\nonumber
L_{2,4,j}=&  -4, 172, 1154, 512, 26 \\
\nonumber
L_{2,5,j}=&  -202, 480, -6775, -12920, -2760, -142 \\
\nonumber
L_{2,6,j}=&  1513, 6840, -6396, 164632, 121926, 17328, 367 \\
\nonumber
L_{2,7,j}=&  13844, -\frac{312669}{2}, \frac{289723}{2}, -537229, -\frac{5764087}{2}, -\frac{2203215}{2}, -119588, 1622 \\
\nonumber
L_{2,8,j}=&  -186286, -659452, 6406508, -11140632, 30368426, 41611544, 10927982, 727168,
16426 \\
\nonumber
L_{2,9,j}=&  -2329677, 29451312, -59944203, -59935254, 168557229, -868313862, -566610531, -112970844, -4278744, -387296 \\
\nonumber
L_{2,10,j}=&  44494564, 197763640, -1758307745, 5510577320, -4801977400, 4018782356, 18152929700, 7911245820, 1036370260, 46448480,
	\\\nonumber&
	 -2514101 
\end{align}
}
\underline{\bf$ L_{4,ij}$}
{\footnotesize
\begin{align}
\nonumber
L_{4,0,j}=&  -2 \\
\nonumber
L_{4,1,j}=&  16, 16 \\
\nonumber
L_{4,2,j}=&  -116, -320, -116 \\
\nonumber
L_{4,3,j}=&  744, 4284, 4512, 820 \\
\nonumber
L_{4,4,j}=&  -4312, -44336, -102280, -55168, -5776 \\
\nonumber
L_{4,5,j}=&  25616, 373920, 1655720, 1986160, 627360, 42536 \\
\nonumber
L_{4,6,j}=&  -195248, -2844360, -20344398, -48520664, -34126080, -6947520, -337928 \\
\nonumber
L_{4,7,j}=&  1664756, 25142754, 199709804, 862903454, 1214720416, 544748526, 77600656, 2821772 \\
\nonumber
L_{4,8,j}=&  -8817028, -271133008, -1911844984, -11428822464, -30753313852, -27284165488, -8379868600, -887874560, -23816660 \\
\nonumber
L_{4,9,j}=&  -3185856, 2178837936, 24460208856, 123319409520, 563440122936, 960048581148, 570029457144, 127532142972, 10362188736,
	\\\nonumber&
	 211088656 \\
\nonumber
L_{4,10,j}=&  -732385616, -3826935440, -314110983830, -1604587100080, -7561982979400, -24271705366600, -27091893699520,
	\\\nonumber&
	-11401711597200, -1944638947880, -123003909760, -2109993836 
\end{align}
}

\subsubsection{$J_1$-$J_d$ model}
First and third neighbor (across the hexagon) model on the kagome lattice (see Eq. \ref{EQ-defJ1J2}).\\
\underline{\bf$ L_{0,ij}$}
{\footnotesize
\begin{align}
\nonumber
L_{0,0,j}=&   \ln2 \\
\nonumber
L_{0,1,j}=&    0,0 \\
\nonumber
L_{0,2,j}=&    \frac{3}{2}, 0, \frac{3}{4} \\
\nonumber
L_{0,3,j}=&    0, 0, 0, \frac{3}{4} \\
\nonumber
L_{0,4,j}=&    -\frac{51}{4}, 9, -\frac{15}{2}, 0, -\frac{15}{8} \\
\nonumber
L_{0,5,j}=&    0, -45, -\frac{15}{2}, -\frac{45}{2}, 0, -\frac{45}{4} \\
\nonumber
L_{0,6,j}=&    \frac{2139}{4}, -\frac{1377}{2}, \frac{2241}{4}, -\frac{615}{2}, \frac{549}{4}, 0, \frac{63}{8} \\
\nonumber
L_{0,7,j}=&    -\frac{483}{2}, 6426, -\frac{3591}{2}, 2919, -\frac{1491}{2}, \frac{2793}{2}, 0, \frac{2751}{8} \\
\nonumber
L_{0,8,j}=&    -\frac{408459}{8}, 97311, -\frac{173109}{2}, 89517, -\frac{87855}{4}, 18306, -\frac{4467}{2}, 0, \frac{12753}{8} \\
\nonumber
L_{0,9,j}=&    77787, -1389825, \frac{2193129}{2}, -\frac{1769283}{2}, 493074, -\frac{666711}{2}, \frac{305523}{2}, -\frac{222993}{2}, 0, -\frac{64545}{4} \\
\nonumber
L_{0,10,j}=&    \frac{34646571}{4}, -\frac{43752975}{2}, \frac{82399395}{4}, -30741990, 10599435, -\frac{20654355}{2}, \frac{1636005}{2}, -\frac{2471625}{2}, -\frac{1312725}{4}, 0, -\frac{1003299}{8} \\
\nonumber
L_{0,11,j}=&    -28988091, \frac{891753291}{2}, -\frac{2317648509}{4}, \frac{1886633727}{4}, -\frac{746930745}{2}, \frac{305988375}{2}, -143379357, \frac{186444885}{4}, -27631923, 10959333,
	\\\nonumber&
	 0, \frac{7095627}{8} 
\end{align}
}
\underline{\bf$ L_{2,ij}$}
{\footnotesize
\begin{align}
\nonumber
L_{2,0,j}=&   1 \\
\nonumber
L_{2,1,j}=&   -2, -1 \\
\nonumber
L_{2,2,j}=&   4, 8, 0 \\
\nonumber
L_{2,3,j}=&   -3, -48, -12, 2 \\
\nonumber
L_{2,4,j}=&   -4, 176, 260, -32, 5 \\
\nonumber
L_{2,5,j}=&   -202, 225, -2935, -270, 60, -21 \\
\nonumber
L_{2,6,j}=&   1513, -4206, 13170, 20210, -4170, 1104, -\frac{399}{2} \\
\nonumber
L_{2,7,j}=&   13844, -74704, 139755, -371574, 13615, -3738, 938, 160 \\
\nonumber
L_{2,8,j}=&   -186286, 1145568, -1664724, 1483992, 2892272, -870832, 248420, -61008, 11421 \\
\nonumber
L_{2,9,j}=&   -2329677, 11526543, -36200385, 64998198, -81429093, 14618331, -3893763, 1180962, -311904, 37370 \\
\nonumber
L_{2,10,j}=&   44494564, -323940580, 753910650, -746061580, 248953155, 673343648, -251041900, 86401210, -21228160, 4486600, -\frac{1698455}{2} \\
\nonumber
L_{2,11,j}=&   568071766, -\frac{5959595279}{2}, \frac{19215984161}{2}, -\frac{49178193933}{2}, 33878328495, -30148742943, 8479066530, -2914976526, \frac{1749626615}{2},
	\\\nonumber&
 -246955709, 58891206, -8569254 
\end{align}
}
\underline{\bf$ L_{4,ij}$}
{\footnotesize
\begin{align}
\nonumber
L_{4,0,j}=&   -2 \\
\nonumber
L_{4,1,j}=&   16, 8 \\
\nonumber
L_{4,2,j}=&   -116, -160, -18 \\
\nonumber
L_{4,3,j}=&   744, 2256, 816, 2 \\
\nonumber
L_{4,4,j}=&   -4312, -25408, -21904, -1664, 80 \\
\nonumber
L_{4,5,j}=&   25616, 230040, 432080, 112440, -5040, 348 \\
\nonumber
L_{4,6,j}=&   -195248, -1677144, -6525636, -4301956, -72768, 8160, -1554 \\
\nonumber
L_{4,7,j}=&   1664756, 12109328, 73372866, 114138990, 20042176, -1978872, 309176, -23372 \\
\nonumber
L_{4,8,j}=&   -8817028, -144204192, -582286200, -2164961352, -1167996400, 52576712, -8761840, 646272, 3726 \\
\nonumber
L_{4,9,j}=&   -3185856, 1911860064, 4580146296, 27401069016, 40740465744, 3935376000, -654091248, 150908760, -24715584, 2009720 \\
\nonumber
L_{4,10,j}=&   -732385616, -8148405280, -100466877420, -185402636680, -927474639540, -417021899020, 47710189400, -9871806500,
	\\\nonumber&
	 1614779960, -186644000, 9926642 \\
\nonumber
L_{4,11,j}=&   31397585488, -188742865432, 1997911450150, 1302805541730, 11994354463608, 19235992257504, 306125884614, -137918907786,
	\\\nonumber&
	60250178296, -14025647636, 2387379192, -205328478 
\end{align}
}

\subsubsection{$J_1$-$J_2$-$J_d$ model}
First, second and third neighbor (across the hexagon) model on the kagome lattice.
The polynomials are
\begin{align}
	L_{2\nu,i,j,k}=\sum_{k=0}^i\sum_{j=0}^{i-k} J_i^{i-j-k}J_2^jJ_d^k
\end{align}
For example, for $i=3$, the coefficients are:  $L_{2\nu,3,0,0}$, $L_{2\nu,3,1,0}$, $L_{2\nu,3,2,0}$, $L_{2\nu,3,3,0}$, $L_{2\nu,3,0,1}$, $L_{2\nu,3,1,1}$, $L_{2\nu,3,2,1}$, $L_{2\nu,3,0,2}$, $L_{2\nu,3,1,2}$, $L_{2\nu,3,0,3}$.\\
\underline{\bf$ L_{0,ij}$}
{\footnotesize
\begin{align}
\nonumber
L_{0,0,j,k}=&  \ln2 \\
\nonumber
L_{0,1,j,k}=&  0,0,0 \\
\nonumber
L_{0,2,j,k}=&  \frac{3}{2}, 0, \frac{3}{2}, 0, 0, \frac{3}{4} \\
\nonumber
L_{0,3,j,k}=&  0, -\frac{9}{2}, 0, 0, 0, -9, 0, 0, 0, \frac{3}{4} \\
\nonumber
L_{0,4,j,k}=&  -\frac{51}{4}, 12, 0, 0, -\frac{51}{4}, 9, 12, 21, 0, -\frac{15}{2}, -6, -\frac{15}{2}, 0, 0, -\frac{15}{8} \\
\nonumber
L_{0,5,j,k}=&  0, 165, -\frac{285}{2}, \frac{165}{2}, 0, 0, -45, 285, -\frac{465}{2}, 300, 0, -\frac{15}{2}, \frac{75}{2}, -\frac{135}{2}, -15, -\frac{45}{2}, 165, -\frac{45}{2}, 0, 0, -\frac{45}{4} \\
\nonumber
L_{0,6,j,k}=&  \frac{2139}{4}, -819, -\frac{1449}{2}, 663, 315, 0, \frac{2139}{4}, -\frac{1377}{2}, 54, -\frac{8739}{2}, 1566, -\frac{2187}{2}, 0, \frac{2241}{4}, 1440, -\frac{5445}{2}, 1800, \frac{1125}{2}, -\frac{615}{2}, -81, -936,
	\\\nonumber&
	 165, \frac{549}{4}, 657, \frac{549}{4}, 0, 0, \frac{63}{8} \\
\nonumber
L_{0,7,j,k}=&  -\frac{483}{2}, -\frac{30681}{2}, 25053, -10962, \frac{5691}{2}, -\frac{44289}{4}, 0, -\frac{483}{2}, 6426, -22260, 45087, -\frac{47649}{2}, 6405, -\frac{67683}{2}, 0, -\frac{3591}{2}, -756, \frac{11571}{2},
	\\\nonumber&
	\frac{39417}{2}, -\frac{8925}{2}, 189, 2919, -\frac{49161}{2}, -6132, -\frac{46683}{2}, -924, -\frac{1491}{2}, -\frac{5985}{4}, \frac{2499}{4}, 1575, \frac{2793}{2}, -\frac{10857}{2}, \frac{2793}{2}, 0, 0, \frac{2751}{8} \\
\nonumber
L_{0,8,j,k}=&  -\frac{408459}{8}, 110334, 168378, -362052, \frac{329469}{2}, -27396, -11184, 0, -\frac{408459}{8}, 97311, -164916, 959034, -1111674, 1005927,
	\\\nonumber&
	 -132474, 192024, 0, -\frac{173109}{2}, -387282, 862149, -1174404, 785022, -230052, -\frac{137319}{2}, 89517, 53982, 718404, -148548, 187728,
	\\\nonumber&
	 -13248, -\frac{87855}{4}, -223434, 416511, -233778, -\frac{163785}{4}, 18306, -2328, 75225, -11682, -\frac{4467}{2}, -61650, -\frac{4467}{2}, 0, 0, \frac{12753}{8} \\
\nonumber
L_{0,9,j,k}=&  77787, 2561598, -\frac{12045861}{2}, 2492136, 1329210, 1721169, -974223, \frac{4060989}{2}, 0, 77787, -1389825, 3166209, -\frac{16866765}{2}, -2009853
	\\\nonumber&
	, \frac{12267315}{2}, 2007180, -1771578, 6139773, 0, \frac{2193129}{2}, -594675, 6741306, -23197428, 17155557, -\frac{19011969}{2}, 1050003, 64827,
	\\\nonumber&
	 -\frac{1769283}{2}, 3837645, 5167827, -8766918, 11837610, 5876253, 450396, 493074, -1772847, \frac{1500957}{2}, -\frac{16493139}{2}, \frac{2952531}{2}, -304101,
		\\\nonumber&
 -\frac{666711}{2}, 1600479, \frac{6957063}{2}, 1406133, 33561, \frac{305523}{2}, \frac{117261}{2}, 311256, -244449, -\frac{222993}{2}, 168885, -\frac{222993}{2}, 0, 0, -\frac{64545}{4} 
\end{align}
}
\underline{\bf$ L_{2,ij}$}
{\footnotesize
\begin{align}
\nonumber
L_{2,0,j,k}=&  1 \\
\nonumber
L_{2,1,j,k}=&  -2, -2, -1 \\
\nonumber
L_{2,2,j,k}=&  4, 16, 4, 8, 8, 0 \\
\nonumber
L_{2,3,j,k}=&  -3, -81, -96, -3, -48, -114, -48, -12, -12, 2 \\
\nonumber
L_{2,4,j,k}=&  -4, 172, 1094, 432, -4, 176, 1108, 1188, 216, 260, 396, 260, -32, -32, 5 \\
\nonumber
L_{2,5,j,k}=&  -202, 605, -5595, -11645, -1360, -202, 225, -7630, -17230, -11430, -680, -2935, -8465, -7970, -3245, -270, -410, -270,
	\\\nonumber&
	 60, 60, -21 \\
\nonumber
L_{2,6,j,k}=&  1513, 4104, -14253, 132562, 98088, 6768, 1513, -4206, 15912, 187998, 251238, 104100, 3384, 13170, 122658, 183249, 132630, 26370,
	\\\nonumber&
	 20210, 40632, 38550, 18080, -4170, -2646, -4170, 1104, 1104, -\frac{399}{2} \\
\nonumber
L_{2,7,j,k}=&  13844, -151620, 139083, -\frac{189371}{2}, -\frac{4759391}{2}, -\frac{1275309}{2}, -106036, 13844, -74704, 299964, -1293383, -3589292, -3827614,
	\\\nonumber&
	-775061, -53018, 139755, -\frac{1966153}{2}, -\frac{6770421}{2}, -3512439, -1991059, -\frac{262773}{2}, -371574, -1154895, -1749132, -990822,
	\\\nonumber&
	 -354109, 13615, -\frac{82789}{2}, -86618, \frac{67319}{2}, -3738, 7651, -3738, 938, 938, 160 \\
\nonumber
L_{2,8,j,k}=&  -186286, -137536, 6324260, -10846320, 16142374, 32058256, 4069692, 764352, -186286, 1145568, 857352, -4756328, 41699268,
	\\\nonumber&
 67176864, 54159980, 3657208, 382176, -1664724, -2088608, 40719726, 82220052, 65417170, 26510312, 811044, 1483992, 19842784,
 	\\\nonumber&
	42915800, 46137688, 18554568, 4060012, 2892272, 6650764, 10425242, 6383884, 1625812, -870832, -1232896, -1281328, -592716,
	\\\nonumber&
	248420, 127572, 248420, -61008, -61008, 11421 \\
\nonumber
L_{2,9,j,k}=&  -2329677, 26960814, -56758545, -115342752, 319743243, -580389138, -314815344, -55709883, 7338168, -2329677, 11526543,
	\\\nonumber&
	 -111054150, 154668708, -220814460, -822646359, -1427828508, -576786222, -27272862, 3669084, -36200385, 128462472,
	\\\nonumber&
	 -152834616, -1365041772, -1793770353, -1226646711, -287537220, -20688669, 64998198, -147995460, -706484844,
 	\\\nonumber&
	-1401175632, -944463996, -380586006, -33586860, -81429093, -244445922, -436794300, -456663087, -172316772,
	\\\nonumber&
	-59040405, 14618331, 3083328, 24409008, -13919346, 9044406, -3893763, 2878929, 7769322, -5141667,1180962, -965718,
	\\\nonumber&
	 1180962, -311904, -311904, 37370 
\end{align}
}
\underline{\bf$ L_{4,ij}$}
{\footnotesize
\begin{align}
\nonumber
L_{4,0,j,k}=&  -2 \\
\nonumber
L_{4,1,j,k}=&  16, 16, 8 \\
\nonumber
L_{4,2,j,k}=&  -116, -320, -116, -160, -160, -18 \\
\nonumber
L_{4,3,j,k}=&  744, 4284, 4512, 744, 2256, 5304, 2256, 816, 816, 2 \\
\nonumber
L_{4,4,j,k}=&  -4312, -44336, -101080, -52416, -4312, -25408, -112016, -115344, -26208, -21904, -45360, -21904, -1664, -1664, 80 \\
\nonumber
L_{4,5,j,k}=&  25616, 370880, 1606560, 1893640, 529280, 25616, 230040, 1816880, 3483320, 2042400, 264640, 432080, 1521760, 1531960, 446440,
	\\\nonumber&
	112440, 210160, 112440, -5040, -5040, 348 \\
\nonumber
L_{4,6,j,k}=&  -195248, -2709360, -19245948, -45281720, -30254952, -4898880, -195248, -1677144, -23323248, -78032388, -86694216,
	\\\nonumber&
	 -31629360, -2449440, -6525636, -38094528, -64060632, -40551744, -7494570, -4301956, -12582120, -12474960, -4348456,
	\\\nonumber&
	 -72768, -139392, -72768, 8160, 8160, -1554 \\
\nonumber
L_{4,7,j,k}=&  1664756, 22589784, 180910170, 774556076, 1076124056, 428284248, 46001648, 1664756, 12109328, 234583524, 1368386446,
	\\\nonumber&
	2616981556, 1883579600, 443305576, 23000824, 73372866, 745015768, 2003634696, 2122598688, 917875952, 108075660, 114138990,
	\\\nonumber&
	 507471972, 776460300, 512949696, 122534342, 20042176, 51422728, 51322348, 19126576, -1978872, -2989952, -1978872, 309176,
	\\\nonumber&
	 309176, -23372 \\
\nonumber
L_{4,8,j,k}=&  -8817028, -254148640, -1625660416, -9686835696, -26227733120, -22350160640, -5601590400, -482074368, -8817028,
	\\\nonumber&
	 -144204192, -1969389360, -18660951416, -60912572640, -75690091560, -37094987200, -5794221968, -241037184,
	\\\nonumber&
	 -582286200, -11246859008, -49514809104, -82150709952, -60903194936, -18467102752, -1405768368, -2164961352,
	\\\nonumber&
	 -15306006176, -34595467888, -35409256832, -16845984288, -2759685248, -1167996400, -4180145600, -6052760560,
	\\\nonumber&
	 -4063750784, -1125824156, 52576712, 95284496, 84481376, 55329600, -8761840, -13587168, -8761840, 646272, 646272, 3726 \\
\nonumber
L_{4,9,j,k}=&  -3185856, 2368385640, 20923820496, 97265440680, 446985312264, 773024430096, 415486343928, 72171633312, 5155399872,
	\\\nonumber&
	-3185856, 1911860064, 21493550160, 196755989184, 1102419923760, 2287787105664, 1979909036496, 673593790032, 73921567680,
	\\\nonumber&
	2577699936, 4580146296, 125301420864, 965794984272, 2518230677256, 2904997704336, 1580660769672, 337698284040,
	\\\nonumber&
	 17974879128, 27401069016, 351660734352, 1196730113328, 1794472817280, 1340425352664, 474314181840, 52563002388,
	\\\nonumber&
	 40740465744, 209933812080, 418462570368, 418254145704, 207522051228, 42963499800, 3935376000, 12935049120, 19258053192,
	\\\nonumber&
	 12803011272, 3037163364, -654091248, -1246816152, -1297966500, -529926528, 150908760, 216561024, 150908760, -24715584,
	\\\nonumber&
	 -24715584, 2009720 
\end{align}
}

\subsection{bcc lattice}
We provide one more term than in  J. Oitmaa and E. Bornilla, Phys. Rev. B p14228 (1996).
{\footnotesize
\begin{align}
\nonumber
	L_{0,i}=&\ln2, 0, 3, -3, \frac{21}{2}, 45, \frac{765}{2}, -\frac{4641}{2}, \frac{437769}{4}, -1042347,
	\frac{80383431}{2}, -\frac{491491737}{2}, \frac{51759195231}{4}, -\frac{449076340089}{2},
	\frac{961339472666851}{64}
\end{align}
}


\begin{thebibliography}{23}
\bibitem [{Note1()}]{SPA}%
\BibitemOpen
$G_h^\PA(e)$ being a rational function of $e$, $s_h^\PA(e)$ is evaluated as a function of $e$ also.
For gapped systems, $G_h^\PA(e)$ is first decomposed in simple fractions and then integrated on $e$.\BibitemShut {NoStop}%
\end{thebibliography}
\end{document}